# Lake Symbols for Island Parsing

Katsumi Okuda[a,b] and Shigeru Chiba[a]

a   The University of Tokyo
b   Mitsubishi Electric Corporation

**Abstract**

**Context**   An island parser reads an input text and builds the parse (or abstract syntax) tree of only the programming constructs of interest in the text. These constructs are called *islands* and the rest of the text is called *water*, which the parser ignores and skips over. Since an island parser does not have to parse all the details of the input, it is often easy to develop but still useful enough for a number of software engineering tools. When a parser generator is used, the developer can implement an island parser by just describing a small number of grammar rules, for example, in Parsing Expression Grammar (PEG).

**Inquiry**   In practice, however, the grammar rules are often complicated since the developer must define the *water* inside the *island*; otherwise, the island parsing will not reduce the total number of grammar rules. When describing the grammar rules for such *water*, the developer must consider other rules and enumerate a set of symbols, which we call *alternative* symbols. Due to this difficulty, island parsing seems to be not widely used today despite its applicability to software engineering tools.

**Approach**   This paper proposes the lake symbols for addressing this difficulty in developing an island parser. It also presents an extension to PEG for supporting the lake symbols. The lake symbols automate the enumeration of the alternative symbols for the *water* inside an *island*. The paper proposes an algorithm for translating the extended PEG to the normal PEG, which can be given to an existing parser generator based on PEG.

**Knowledge**   The user can use lake symbols to define *water* without specifying each *alternative* symbol. Our algorithms can calculate all *alternative* symbols for a lake symbol, based on where the lake symbol is used in the grammar.

**Grounding**   We implemented a parser generator accepting our extended PEG and implemented 36 island parsers for Java and 20 island parsers for Python. Our experiments show that the lake symbols reduce 42 % of grammar rules for Java and 89 % of rules for Python on average, excluding the case where islands are expressions.

**Importance**   This work simplifies the description of grammars for island parsing. Lake symbols enable the user to define the *water* inside the *island* simpler than before. Defining *water* inside the *island* is essential to apply island parsing for practical programming languages.

## The Art, Science, and Engineering of Programming

## 1 Introduction

Island parsing is a promising technique for the development of software engineering tools. Island parsers extract only programming constructs of interest as *islands* from an input text and they ignore the rest of the text as *water*. Such *incomplete* parsers are often easy to develop but still useful enough for a number of software engineering tools. These tools do not need the complete parse tree or the abstract syntax tree (AST) of the input program. They only need a parse tree or an AST for the interesting parts of the program. Suppose that we develop a tool for measuring the chromatics complexity [17] of programs. The tool only needs a parse tree for all the if/else statements and other control-flow statements. We can just ignore the rest of the program. An island parser fits this application; the statements like if/else are *islands* while the rest is *water*. An island parser is also suitable for source model extraction from incomplete program [18], multilingual parsing [1, 24, 26], extracting programming constructs from documentations [3, 22], and lightweight impact analysis [19].

The grammar for an island parser is described in a formal language such as PEG (Parsing Expression Grammar) [9], SDF (Syntax Definition Formalism) [12], or TXL [5]. This grammar consists of the rules for islands and water. Describing this grammar is often easier than describing the corresponding full-featured grammar, which is for parsing a whole program. The rules for the water are usually simple, in an ideal case, just one wildcard character, and thus the grammar for island parsing consists of only a small number of rules. The island parser can be constructed from this grammar in the same way as ordinal parsers.

In practice, the grammar for island parsing consists of a small number of rules only when most parts of a program are recognized as water. To do so, some inner parts of the island must be also water, which is not fully parsed. For example, if an island is a if/else statement, the expressions included in that statement should be recognized as water. However, the rules for such water in the middle of an island are complicated and difficult to describe. Due to this difficulty, island parsing seems to be not widely used today despite its applicability to software engineering tools.

In this paper, we propose the lake symbols to mitigate the difficulty in describing the rules for the water in the middle of an island. We call this water a *lake*. A lake symbol is a special symbol used in the grammar to represent a lake in the grammar. In the simplest case, the lake symbol works as a special wildcard that matches any character except the parts of the island. The parser needs to know what text patterns must not be taken as water to exclude the parts of the island from the lake. These patterns are given by terminal or non-terminal symbols, which we call the *alternative symbols*. The lake symbol automates the enumeration of the alternative symbols so that it can be used as a simple wildcard-like symbol. The user can also specify inner programming constructs in the lake by writing the rule for the lake symbol. This feature enables the handling of nested islands inside the lake. We also propose an extension to PEG for supporting our lake symbols. We present an algorithm to translate this extended PEG into the normal PEG, which can be used as an input to an existing PEG-based parser or parser-generator.

Furthermore, we implemented *PEGIsland*, an island-parser generator supporting our extended PEG. By using PEGIsland, we implemented 36 island parsers for Java programs and 20 ones for Python programs. We compared the number of their grammar rules with the number of the rules for the full-featured Java/Python parser. The comparison revealed that our extended PEG effectively reduces the number of the rules.

The contribution of this paper is twofold:

- We propose the lake symbols and an extended PEG supporting the lake symbols. It mitigates the difficulty in describing the grammar rule for the water in the middle of an island. This contributes to reducing the total number of rules for an island parser and thereby improving the usefulness of the island parser.
- We implemented island parsers based on the extended PEG and revealed that the extended PEG effectively reduced the size of the grammar for island parsing.

In the rest of this paper, we first present island parsing and our motivating problem. We then propose the lake symbols, an extension to PEG for supporting the lake symbols, and the algorithm for translating the extended PEG into the normal PEG. We also present our experiments and related work. Finally, we conclude this paper.

## 2 Motivating Example

A parsing expression grammar (PEG) [9] is one of the formal grammars that are used for building a top-down parser (see appendix A for details on PEGs). Listing 1 shows an example of PEG. It specifies a simple language; its program consists of statements (stmt) and function definitions (func_def). A statement is either a block (block), an if/else statement (if_else_stmt), or an expression statement (exp_stmt). An expression statement is an expression (expr) with a semicolon at the end. An expression supports several binary operators such as + and > and its terms are parenthesized expressions, function calls, or number literals. Note that the terms may be a lambda expression (lambda_expr), which includes a block and the statements in that block.

The starting symbol of this grammar is program. The grammar is scanner-less and its terminal symbols (or lexical tokens) are every character of its programs. spacing is a nonterminal symbol used for recognizing whitespace and skipping it. For example, the parsing expression for program starts with spacing for skipping the whitespace at the beginning of the program.

A program in listing 2 is an example written in the language specified by listing 1. Lines 1 to 21 are a function definition recognized by the nonterminal symbol func_def. Lines 3 to 19 are an if/else statement recognized by if_else_stmt. Lines 4 to 15 are a lambda expression that includes another if/else statement on lines 5 to 13. Lines 8 to 13 are also an if/else statement that is the *else* part of the if/else statement on lines 5 to 13.

Figure 1 is the parse tree obtained after the parser based on the grammar in listing 1 parses the program in listing 2. Each node of the tree is labeled with a nonterminal

**Lake Symbols for Island Parsing**

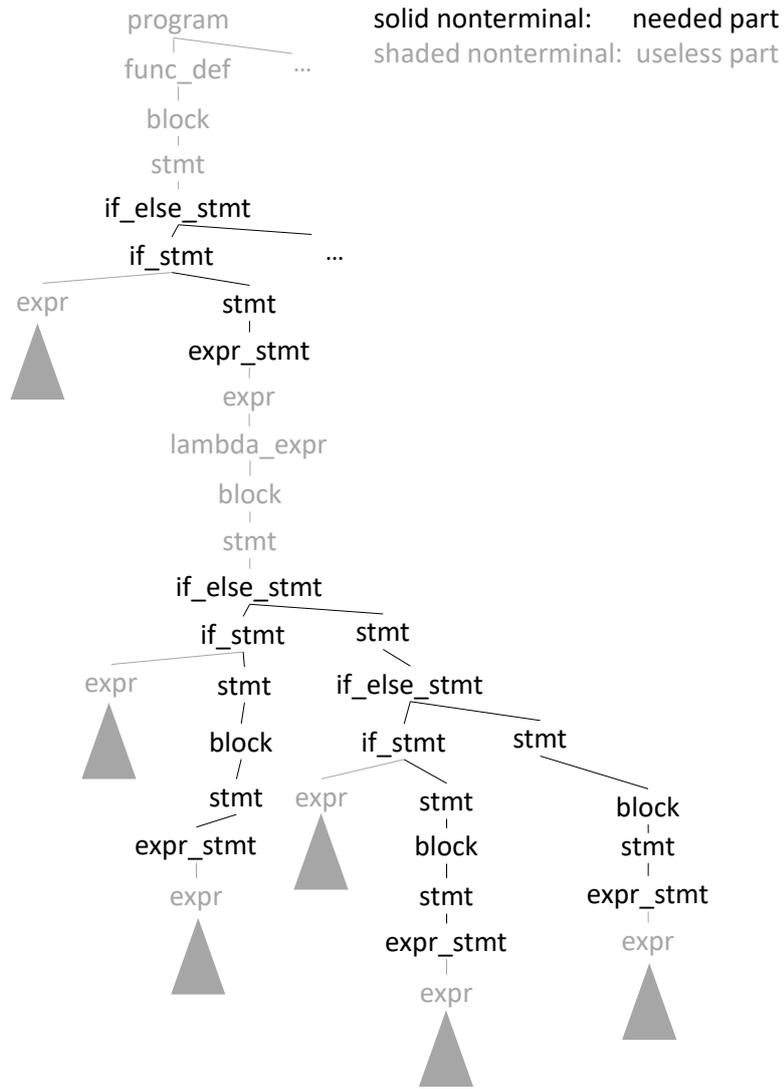

**Figure 1** The parse tree for the program in listing 2

▬ **Listing 1** The PEG for a simple language

```
1  program <- spacing (stmt / func_def)*
2  func_def <- FUNCTION ID LPAREN ID? RPAREN block
3  if_else_stmt <- if_stmt (ELSE stmt)?
4  if_stmt <- IF LPAREN expr RPAREN stmt
5  stmt <- block / if_else_stmt / exp_stmt
6  exp_stmt <- expr SEMICOLON
7  block <- LBRACE stmt* RBRACE
8  expr <- rel_expr (ASSIGN rel_expr)*
9  rel_expr <- add_expr ((EQ / GT / LT) add_expr)*
10 add_expr <- mul_expr ((PLUS / MINUS) mul_expr)*
11 mul_expr <- pri_expr ((MUL / DIV) pri_expr)*
12 pri_expr <- LPAREN expr RPAREN / lambda_expr / funcall / NUMBER / STRING / ID
13 lambda_expr <- LAMBDA LPAREN ID? RPAREN block
14 funcall <- ID LPAREN expr? RPAREN
15
16 spacing <- [ \t\n]*
17 ID <- [_a-z]+ spacing
18 LPAREN <- '(' spacing
19 RPAREN <- ')' spacing
20 LBRACE <- '{' spacing
21 RBRACE <- '}' spacing
22 SEMICOLON <- ';' spacing
23 IF <- 'if' spacing
24 ELSE <- 'else' spacing
25 STRING <- '"' [^"]* '"' spacing
26 FUNCTION <- 'function' spacing
27 LAMBDA <- 'lambda' spacing
28 NUMBER <- [0-9]+ spacing
29 ASSIGN <- '=' spacing
30 EQ <- '==' spacing
31 GT <- '>' spacing
32 LT <- '<' spacing
33 PLUS <- '+' spacing
34 MINUS <- '-' spacing
35 MUL <- '*' spacing
36 DIV <- '/' spacing
```

symbol. For example, the root of the tree is labeled with program. The details of the subtree labeled with expr is omitted and represented by a gray triangle in this figure.

Island parsing is a technique for recognizing only interesting program constructs such as an if/else statement in the given program. The benefit of island parsing is that the number of the rules in the grammar is reduced since we can eliminate the rules for the uninteresting constructs from the grammar. The program text of these uninteresting constructs is skipped by using a wildcard-like symbol. The interesting program constructs are called *islands* and recognized by the parser while the rest of the constructs are called *water* and skipped by the parser. Islands and water are also called *sea*.

**Lake Symbols for Island Parsing**

■ **Listing 2**  A program in the language in listing 1

```
1  function make_cmp_f(x)
2  {
3      if (x == 0)
4          func = lambda (y) {
5              if (y > x) {
6                  result = "positive";
7              }
8              else if (y < x) {
9                  result = "negative";
10             }
11             else {
12                 result = "zero";
13             }
14             result;
15         };
16     else
17         func = lambda (y) {
18             ...
19         };
20     func;
21 }
22
23 compare = make_cmp_f(0);
24 compare(1);
```

Suppose that we are interested in only the if/else statements for the language in listing 1 when we are developing a simple tool for that language. We assume that the tool rewrites if/else statements so that they will be written with curly braces. Coding conventions such as MISRA C [2] state that the control structures should be written with curly braces. In the case of the listing 2, the tool rewrites the outermost if/else statement on lines 3 to 19 since it does not include braces for the if clause and the else clause. The parse tree that we would want to obtain does not need to contain the nodes or leaves that are irrelevant to the if/else statements. It would be the tree consisting of only the nodes labeled with a solid name in figure 1. The gray nodes or the gray triangles would not be included in the tree.

An island parser that we need recognizes only an if/else statement as an *island* but skips the others as *water*. The grammar for that parser does not need to include the rule for func_def. It is obtained by eliminating the rule for func_def from listing 1 and adding the following three rules:

```
1  program <- spacing program_sea*
2  program_sea <- if_else_stmt / program_water
3  program_water <- STRING / .
```

We introduce two new nonterminal symbols program_sea and program_water. A program is the repetition of program_sea, which is either if_else_stmt (i.e. an island) or program_water. program_water is a wildcard-like symbol. It recognizes either a string literal or any character matching a wildcard character ".". Note that the prioritized

■ **Listing 3** The grammar for the island parser

```
1  program <- spacing program_sea*
2  program_sea <- if_else_stmt / program_water
3  program_water <- STRING / .
4  if_else_stmt <- if_stmt (ELSE stmt)?
5  if_stmt <- IF LPAREN expr RPAREN stmt
6  stmt <- block / if_else_stmt / exp_stmt
7  exp_stmt <- expr SEMICOLON
8  block <- LBRACE stmt* RBRACE
9  expr <- expr_sea+
10 expr_sea <- if_else_stmt / expr_water
11 expr_water <- LPAREN expr_sea* RPAREN / block / STRING / !(SEMICOLON / RPAREN / RBRACE) .
12
13 spacing <- [ \t\n]*
14 LPAREN <- '(' spacing
15 RPAREN <- ')' spacing
16 LBRACE <- '{' spacing
17 RBRACE <- '}' spacing
18 SEMICOLON <- ';' spacing
19 IF <- 'if' spacing
20 ELSE <- 'else' spacing
21 STRING <- '"' [^"]* '"' spacing
```

choice operator / gives a higher priority to if_else_stmt than program_water. Otherwise, program_sea would not recognize an if_else_stmt at all. The rule for program_water is not

```
program_water <- .
```

since this would let the parser recognize a string literal like "if(x)y;" wrongly as if_else_stmt.

Next, we remove expr from the grammar as well as func_def since the details of expr are not interesting. The grammar without expr and func_def is listed in listing 3. The size of this grammar is approximately only two thirds of the original one in listing 1. In this grammar, the original rule for expr is replaced with a simpler one. It is the repetition of expr_sea, and expr_sea is either if_else_stmt or expr_water as program_sea is.

Note that expr_water is not the same as program_water although both are *water* symbols, which are used as a wildcard matching any symbol.

```
expr_water <- LPAREN expr_sea* RPAREN / block / STRING / !(SEMICOLON / RPAREN / RBRACE) .
```

First, recall that an expression may be a lambda expression, which may include an if/else statement in its body. Thus, the parser must recognize an if/else statement as an island when it is included in the expression. expr_water must be aware of a parenthesized expression, a block, and a string literal since they may include an if/else statement. On the other hand, program_water must be aware of only a string literal.

Second, the last option of the prioritized choice / in expr_water is not only a wildcard character . but:

```
!(SEMICOLON / RPAREN / RBRACE) .
```

The wildcard character follows a *not* predicate (negative lookahead). This reads as matching any character except a semicolon, a right parenthesis, and a right brace. Since they are parts of an island, they should not be recognized as a part of water. They are sentinels for terminating the repetition of a wildcard. We call these symbols as *alternative symbols* in this paper. Note that the wildcard character in the rule for program_water does not follow a *not* predicate.

Since expr appears in the parsing expressions (the right-hand side of <- for the rule) for other nonterminals such as if_stmt and exp_stmt, its water symbol expr_water may not match the symbols following expr in those parsing expressions. These symbols are alternative symbols for expr_water.

For example, the parsing expression for exp_stmt is:

```
expr SEMICOLON
```

Since expr matches the repetition of if_else_stmt or expr_water, expr_water may not match a semicolon. Note that the PEG-based parser can be regarded as the parser always finding the longest match. Unless a semicolon is excluded from the symbols that expr_water matches, a semicolon would be absorbed into expr and thus not recognized as the last token of expr_stmt.

The alternative symbols for expr_water are not only the symbols following expr. They are also the symbols that may appear at the grammatical position where expr_water may appear. In other words, the alternative symbols include the symbol that the parser must consume as part of a different non-terminal when the parser can consume it as expr_water. Thus, the idea of the alternative symbols are different from the follow set used in the LALR(1) parsing.

An example of such an alternative symbol is RBRACE for expr_water. Note that RBRACE does not follow expr. However, RBRACE is included in the alternative symbols for expr_water. The parser must consume it as a part of block when the parser can consume it as expr_water. Suppose that the parser has read the following character/token sequence:

```
if (x < 0) { x = 1;
```

The parser recognizes this as

```
IF LPAREN expr RPAREN LBRACE stmt
```

The parser will next attempt to recognize another occurrence of stmt. It will try block, if_else_stmt, and finally expr_stmt. If the next token is }, that is, the parser reads:

```
if (x < 0) { x = 1; }
```

then the parser does not recognize } (RBRACE) as the first token for block or if_else_stmt. So the parser next attempts expr_stmt but this attempt must fail. The parser must fail to recognize } as expr_water. It must recognize } as the last token for block instead of expr_water. This is why RBRACE is included in the alternative symbols for expr_water.

Island parsing is a useful technique when we only need a parser that recognizes only interesting program constructs, such as an if/else statement, in the given program.

The size of the grammar for island parsing is often smaller than that of the normal grammar. This fact reduces the development cost of the parser. Furthermore, an island parser can recognize program constructs even when a syntax error is included in the program but in the program constructs that the parser does not recognize but it skips.

However, implementing the grammar definition for island parsing is not easy. As we showed above, the rules for *water* symbols tend to be complicated and error-prone. For example, selecting a right set of the alternative symbols is not easy. This difficulty is a burden to the developers when implementing an island parser.

## 3 Lake Symbol

To reduce the development cost of island parsers, we propose an extension to Parsing Expression Grammar (PEG). It supports island parsing by a new kind of symbol named a *lake symbol*.

An island parser parses a program as the repetition of *sea*. *sea* is either an *island* or *water*.

```
program <- sea*
sea <- island / water
```

An island is a symbol representing the program constructs that we want to parse while water is a symbol representing the other program constructs. Since we are not interested in the latter constructs or want to parse them, the naive definition of the water is a wildcard that anything matches.

As we have seen in the previous section, the definition of the water becomes complicated when we are not interested in inner parts of the island. For example, when we were not interested in the expressions included in an if/else statement, we had to define the water that the expressions match and that definition was complicated.

We call such uninteresting inner parts *lakes* because they are *water* in an *island*. As the *sea* consists of islands and water, the *lake* consists of islands and water. Thus, there may be an island in the middle of the lake. The lake is a kind of the sea but it is an inner part of an island. Our lake symbol helps us define the lake. In particular, it automates the calculation of the alternate symbols.

In our extended PEG, a symbol enclosed in angle brackets <> is a lake symbol. Suppose that a string literal STRING is an island. A lake symbol is used as follows:

```
STRING <- '"' <STRING_lake>* '"' spacing
```

Here, <STRING_lake> is a lake symbol. It represents a lake in the string literal. We do not have to write the rule for defining this lake symbol. The lake symbol can be regarded as a wildcard character that any character except " matches. So the parsing expression <STRING_lake>* matches any text enclosed in double quotes ". The excluded character " is automatically detected; this is a difference from the wildcard for the regular expressions.

A lake symbol is not a simple wildcard. We can specify islands in the lake. Suppose that a block is an island.

**Lake Symbols for Island Parsing**

```
block <- '{' <block_lake>* '}'
<block_lake> <- block
```

Here, <block_lake> is a lake symbol. The second line specifies that a block enclosed in another block is an island and thus it is parsed. If there are multiple kinds of islands in the lake, they are enumerated by using the prioritized choice operator /.

The rule for a lake symbol can be recursive. For example,

```
block <- '{' <block_lake>* '}'
<block_lake> <- '{' <block_lake>* '}'
```

In the second rule, <block_lake> is recursively referred to in the parsing expression (the right hand side of ->). This reads as the parser considers the nested curly braces but it does not parse an inner block as an island. It parses only the outermost block as an island. This rule is necessary to identify which closing brace is balanced to an opening brace.

In our extended PEG, water is a special symbol. The parsing expression for water is appended to the parsing expressions for all the *lake* symbols. For example, if the rule for water is:

```
water <- COMMENT / STRING
```

then the following rule for a lake:

```
<block_lake> <- block
```

is treated as an equivalent of this:

```
<block_lake> <- block / COMMENT / STRING
```

Note that water is also used for the top-level rule program. Listing 4 shows the grammar definition written with lake symbols expr_lake and STRING_lake. It is equivalent to listing 3 except the use of lake symbols. water is used in line 2 and defined in line 3 in that grammar definition.

We can generate an island parser from our extended PEG. We first translate the extended PEG into a normal PEG and then use this normal PEG to generate a parser based on packrat parsing [8] or its variants [7]. In the following subsections, we present how our extended PEG is translated into a normal PEG.

### 3.1 Translation into a Normal PEG

Our extended PEG with lake symbols can be represented by a tuple $G = (V_N, V_L, V_T, R, e_s, e_w)$, where $V_N$ is a set of nonterminal symbols, $V_L$ is a set of lake symbols, $V_T$ is a set of terminal symbols, $R$ is a set of rules, $e_s$ is a starting expression, $e_w$ is a global water expression, $V_N \cap V_L = \emptyset$, $V_N \cap V_T = \emptyset$, $V_L \cap V_T = \emptyset$. The extended elements of the tuple is $V_L$ and $e_w$. $V_L$ is a set of symbols enclosed in angle brackets <> and $e_w$ is given by the rule for the special symbol water. If water is not explicitly given in the grammar, $e_w$ is the regular expression !. (i.e. nothing matches $e_w$).

We translate an extended PEG $G$ into a normal PEG $G''$. Here, $G'' = (V'_N, V_T, R'', e_s)$, such that $V'_N = V_N \cup V_L$. The translation is divided into two steps. In the following, $X$,

■ **Listing 4** The grammar written with lake symbols

```
1  program <- spacing program_sea*
2  program_sea <- if_else_stmt / water
3  water <- STRING / .
4  if_else_stmt <- if_stmt (ELSE stmt)?
5  if_stmt <- IF LPAREN expr RPAREN stmt
6  stmt <- block / if_else_stmt / exp_stmt
7  exp_stmt <- expr SEMICOLON
8  block <- LBRACE stmt* RBRACE
9  expr <- <expr_lake>*
10 <expr_lake> <- if_else_stmt / LPAREN <expr_lake>* RPAREN / block
11
12 spacing <- [ \t\n]*
13 LPAREN <- '(' spacing
14 RPAREN <- ')' spacing
15 LBRACE <- '{' spacing
16 RBRACE <- '}' spacing
17 SEMICOLON <- ';' spacing
18 IF <- 'if' spacing
19 ELSE <- 'else' spacing
20 STRING <- '"' <STRING_lake>* '"' spacing
```

$<X>$ and $e$ are meta-variables. $X$ ranges over nonterminal symbols $V_N$, $<X>$ ranges over lake symbols $V_L$, and $e$ ranges over a parsing expression.

**Step 1**

We translate an extended PEG $G$ into an intermediate PEG $G' = (V_N, V_L, V_T, R', e_s)$. In $G'$, the global water expression $e_w$ is appended to the parsing expression of the rule for every lake symbol in $V_L$.

For each rule in $R$, if the rule is $X \leftarrow e$, then $R'$ includes the rule $X \leftarrow e$ as it is. If the rule is $<X> \leftarrow e$, then $R'$ includes the following rule:

$$<X> \leftarrow e/e_w$$

For each lake symbol $<X>$ in $V_L$, if $R$ does not include a rule for $<X>$, $<X> \leftarrow e$, then $R'$ includes the following rule:

$$<X> \leftarrow e_w$$

**Step 2**

For each rule in $R'$, if the rule is $X \leftarrow e$, then $R''$ includes the rule $X \leftarrow e$ as it is. If the rule is $<X> \leftarrow e$, then $R''$ includes the following rule:

$$<X> \leftarrow e/!(s_1/.../s_n) \;.$$

such that $s_i \in V_N \cup V_T$ and $s_1, s_2, ..., s_n$ are the elements of $ALT(e)$, the set of the alternative symbols for $e$, which is the definition of $<X>$ in $R'$. This rule reads as $<X>$ is $e$ or any character that is not the first part of the text string recognized by $s_1, s_2, ..., s_n$. Note that the last period in the rule is a wildcard character.

**Lake Symbols for Island Parsing**

■ **Listing 5** The syntax of parsing expressions
```
1  parsing_expr ::= parsing_expr parsing_expr | parsing_expr '/' parsing_expr
2       | parsing_expr '*' | parsing_expr '+' | parsing_expr '?'
3       | '&' parsing_expr | '!' parsing_expr
4       | grammar_symbol
```

$$
\begin{array}{ll}
\text{block} & \texttt{<- } \underbrace{\underbrace{\underbrace{\underbrace{\text{'\{'}}_{e_1} \underbrace{\text{stmt*}}_{\underbrace{e_2}_{e_3}} \underbrace{\text{'\}'}}_{e_4}}_{e_5}}_{e_6}} \\[2ex]
\text{stmt} & \texttt{<- } \underbrace{\underbrace{\text{expr\_stmt}}_{e_7} \text{ / } \underbrace{\text{block}}_{e_8}}_{e_9} \\[2ex]
\text{expr\_stmt} & \texttt{<- } \underbrace{\underbrace{\underbrace{\text{<elake>*}}_{e_{10}} \underbrace{\text{';'}}_{e_{11}}}_{e_{12}}}_{e_{13}} \\[2ex]
\text{<elake>} & \texttt{<- } \underbrace{\underbrace{\text{! } \underbrace{\text{.}}_{e_{14}}}_{e_{15}}}{}
\end{array}
$$

■ **Figure 2** The parsing expressions in the grammar

### 3.2 Alternative Symbols

*ALT(e)* is the set of the alternative symbols for *e*. The elements of *ALT(e)* are grammar symbols, which are either a terminal, nonterminal, or lake symbol.

We below use meta-variables $e_i$, $e_j$, ... ranging over the parsing expressions in $R'$. $e_i$, $e_j$, ... are location-aware. Location-aware means that two lexically equivalent parsing expressions $e_i$ and $e_j$ are not identical when they belong to a different rule or they are different sub-expressions in the same expression. A parsing expression parsing_expr is defined as in listing 5. We assume that the sequence operator and the prioritized choice operator have left associativity.

Suppose that we have a grammar written in our extended PEG:

```
1  block <- '{' stmt* '}'
2  stmt <- expr_stmt / block
3  expr_stmt <- <elake>* ';'
```

The step 1 presented in section 3.1 translates this PEG into the intermediate PEG shown in figure 2. All the parsing expressions included in this intermediate PEG are also presented. There are 15 parsing expressions. We label 1 to 15 for each parsing expression. The domain of *ALT* for this grammar is these parsing expressions $e_1$ to $e_{15}$.

**Definition of *ALT***

We define $ALT(e_i)$ as the set of the uppermost symbols that the parser may recognize immediately after it fails to recognize $e_i$. When the parser recognizes a non-terminal symbol, it also recognizes its lower-level symbols. We discard these lower-level symbols when calculating $ALT(e_i)$. We use the fixed-point iteration algorithm shown in algorithm 1 to calculate $ALT(e_i)$ for all $e_i$ in the intermediate PEG. The constraints satisfied by the fixed-point is shown in appendix B.1. The input to algorithm 1 is the

---

**Algorithm 1** Calculation of $ALT(e_i)$

---

**Input:** a set of all the parsing expressions $E$ included in the rules $R'$
**Output:** $ALT(e_i)$ includes the alternative symbols for the parsing expression $e_i$

1: **for all** parsing expressions $e_i$ in $E$ **do**
2:     $ALT(e_i) \leftarrow \emptyset$
3: **while** $ALT(e_i)$ is changing for some $e_i$ in $E$ **do**
4:     **for all** parsing expressions $e_i$ in $E$ **do**
5:         **if** $e_i$ is a terminal symbol **then**
6:             do nothing
7:         **else if** $e_i$ is a nonterminal or lake symbol $\mathscr{S}$ and $\mathscr{S} \leftarrow e_j \in R'$ **then**
8:             $ALT(e_j) \leftarrow ALT(e_j) \cup ALT(e_i)$
9:         **else if** $e_i$ is $e_j*$, $e_j+$, or $e_j?$ **then**
10:            $ALT(e_j) \leftarrow ALT(e_i) \cup SUCCEED(e_i)$
11:         **else if** $e_i$ is $!e_j$ **then**
12:            $ALT(e_j) \leftarrow SUCCEED(e_i)$
13:         **else if** $e_i$ is $\&e_j$ **then**
14:            $ALT(e_j) \leftarrow ALT(e_i)$
15:         **else if** $e_i$ is $e_j/e_k$ **then**
16:            $ALT(e_k) \leftarrow ALT(e_i)$
17:            **if** $\epsilon \in BEGINNING(e_k)$ **then**
18:                $ALT(e_j) \leftarrow ALT(e_i) \cup (BEGINNING(e_k) - \{\epsilon\}) \cup SUCCEED(e_k)$
19:            **else**
20:                $ALT(e_j) \leftarrow ALT(e_i) \cup BEGINNING(e_k)$
21:         **else if** $e_i$ is $e_j e_k$ **then**
22:            $ALT(e_j) \leftarrow ALT(e_i)$
23:            **if** $\epsilon \in BEGINNING(e_j)$ **then**
24:                $ALT(e_k) \leftarrow ALT(e_i)$
25:            **else**
26:                $ALT(e_k) \leftarrow \emptyset$

---

rules $R'$ of the intermediate PEG $G'$ obtained at the step 1. Algorithm 1 also takes a set $E$ as the input. $E$ contains all the parsing expressions appearing in the rules $R'$.

The fixed-point iteration terminates after a finite number of iterations since the size of every $ALT(e_i)$ grows monotonically but it is smaller than $|V_N| + |V_L| + |V_T|$. In the worst case, every iteration adds only one element to only one instance of $ALT$. This case may cause the maximum number of iteration, which is still lower than $|E| \cdot (|V_N| + |V_L| + |V_T|)$.

Algorithm 1 uses two functions $BEGINNING(e_i)$ and $SUCCEED(e_i)$. $BEGINNING$ and $SUCCEED$ are similar to $FIRST$ and $FOLLOW$ for context-free grammars. The difference is that $BEGINNING$ and $SUCCEED$ contain not only terminal symbols but also nonterminal symbols. This difference is essential when a grammar is written in PEG, in which a terminal symbol is just a letter. $BEGINNING(e_i)$ is the set of grammar symbols that the parser may first recognize when it starts recognizing the parsing expression $e_i$. When $e_i$ recognizes an empty input, $BEGINNING(e_i)$ includes $\epsilon$. $SUCCEED(e_i)$ is a

---

**Algorithm 2** Calculation of $BEGINNING(e_i)$

---

**Input:** a set of all the parsing expressions $E$ included in the rules $R'$
**Output:** $BEGINNING(e_i)$ includes the grammar symbols that the parser may first recognize when it starts recognizing $e_i$.

1: **for all** parsing expressions $e_i$ in $E$ **do**
2:     $BEGINNING(e_i) \leftarrow \emptyset$
3: **while** $BEGINNING(e_i)$ is changing for some $e_i$ in $E$ **do**
4:     **for all** parsing expressions $e_i$ in $E$ **do**
5:         **if** $e_i$ is a terminal symbol $\alpha$ **then**
6:             $BEGINNING(e_i) \leftarrow \{\alpha\}$
7:         **else if** $e_i$ is a nonterminal or lake symbol $\mathscr{S}$ and $\mathscr{S} \leftarrow e_j \in R'$ **then**
8:             $BEGINNING(e_i) \leftarrow \{\mathscr{S}\}$
9:         **else if** $e_i$ is $e_j?$ or $e_j*$ **then**
10:             $BEGINNING(e_i) \leftarrow BEGINNING(e_j) \cup \{\epsilon\}$
11:         **else if** $e_i$ is $e_j+$ **then**
12:             $BEGINNING(e_i) \leftarrow BEGINNING(e_j)$
13:         **else if** $e_i$ is $!ej$ or $\&e_j$ **then**
14:             $BEGINNING(e_i) \leftarrow \{\epsilon\}$
15:         **else if** $e_i$ is $e_j/e_k$ **then**
16:             $BEGINNING(e_i) \leftarrow BEGINNING(e_j) \cup BEGINNING(e_k)$
17:         **else if** $e_i$ is $e_j e_k$ **then**
18:             **if** $\epsilon \in BEGINNING(e_j)$ **then**
19:                 $BEGINNING(e_i) \leftarrow (BEGINNING(e_j) - \{\epsilon\}) \cup BEGINNING(e_k)$
20:             **else**
21:                 $BEGINNING(e_i) \leftarrow BEGINNING(e_j)$

---

set of grammar symbols that the parser may next recognize after it recognizes $e_i$. The two functions are calculated by algorithm 2 and algorithm 3, respectively. They use fixed-point iteration. Each fixed-point satisfies the constraints shown in appendix B.2 appendix B.3.

The algorithm calculates $ALT(e_i)$ for every parsing expression $e_i$ in $E$, which is the set of all the parsing expressions included in the rules $R'$. For every $e_i$, the algorithm first sets $ALT(e_i)$ to an empty set. Then the algorithm incrementally updates $ALT(e_i)$ until $ALT(e_i)$ does not change for any $e_i$ in $E$. The update is performed according to line 5 to 26 for every $e_i$ in $E$. For example, when $e_i$ is $e_2$ in figure 2, $ALT(e_9)$ is updated to be $ALT(e_9) \cup ALT(e_2)$ according to line 8. Since $e_2$ is a nonterminal symbol stmt and $R'$ includes the rule stmt <- $e_9$, $ALT(e_2)$ is added to $ALT(e_9)$. When $e_i$ is $e_9$, $ALT(e_8)$ is updated to be $ALT(e_9)$ according to line 16. $ALT(e_7)$ is updated to be $ALT(e_7) \cup BEGINNING(e_8)$ according to line 20 because $BEGINNING(e_8)$ does not include $\epsilon$; $e_8$ does not recognize an empty input.

---

**Algorithm 3** Calculation of *SUCCEED*($e_i$)

---
**Input:** a set of all the parsing expressions $E$ included in the rules $R'$
**Output:** *SUCCEED*($e_i$) includes the grammar symbols that the parser may next recognize after $e_i$.
1: **for all** parsing expressions $e_i$ in $E$ **do**
2: $\quad$ *SUCCEED*($e_i$) ← ∅
3: **while** *SUCCEED*($e_i$) is changing for some $e_i$ in $E$ **do**
4: $\quad$ **for all** parsing expressions $e_i$ in $E$ **do**
5: $\quad\quad$ **if** $e_i$ is a terminal symbol **then**
6: $\quad\quad\quad$ do nothing
7: $\quad\quad$ **else if** $e_i$ is a nonterminal or lake symbol $\mathscr{S}$ and $\mathscr{S} \leftarrow e_j \in R'$ **then**
8: $\quad\quad\quad$ *SUCCEED*($e_j$) ← *SUCCEED*($e_i$) ∪ *SUCCEED*($e_j$)
9: $\quad\quad$ **else if** $e_i$ is $e_j$? **then**
10: $\quad\quad\quad$ *SUCCEED*($e_j$) ← *SUCCEED*($e_i$)
11: $\quad\quad$ **else if** $e_i$ is $e_j*$ or $e_j+$ **then**
12: $\quad\quad\quad$ *SUCCEED*($e_j$) ← *SUCCEED*($e_i$) ∪ *BEGINNING*($e_i$) − {$\epsilon$}
13: $\quad\quad$ **else if** $e_i$ is !$e_j$ or &$e_j$ **then**
14: $\quad\quad\quad$ *SUCCEED*($e_j$) ← ∅
15: $\quad\quad$ **else if** $e_i$ is $e_j/e_k$ **then**
16: $\quad\quad\quad$ *SUCCEED*($e_j$) ← *SUCCEED*($e_i$)
17: $\quad\quad\quad$ *SUCCEED*($e_k$) ← *SUCCEED*($e_i$)
18: $\quad\quad$ **else if** $e_i$ is $e_je_k$ **then**
19: $\quad\quad\quad$ *SUCCEED*($e_k$) ← *SUCCEED*($e_i$)
20: $\quad\quad\quad$ **if** $\epsilon \in$ *BEGINNING*($e_k$) **then**
21: $\quad\quad\quad\quad$ *SUCCEED*($e_j$) ← (*BEGINNING*($e_k$) − {$\epsilon$}) ∪ *SUCCEED*($e_k$)
22: $\quad\quad\quad$ **else**
23: $\quad\quad\quad\quad$ *SUCCEED*($e_j$) ← *BEGINNING*($e_k$)

---

### 3.3 Example

In this subsection, we show the calculation of *ALT* for the parsing expressions in figure 2. Before doing so, we first show that block, ';', and '}' are the alternative symbols for <elake>, which is *ALT*($e_{15}$) by using a railroad diagram. The railroad diagram is useful to visualize alternative symbols.

Figure 3 is the railroad diagram for the nonterminal symbol block. stmt in the parsing expression for block has been expanded into its definition. expr_stmt has been also expanded. In figure 3, round boxes depict terminal symbols while square boxes depict nonterminal or lake symbols.

The alternative symbols for <elake> are what the parser may recognize when it can next recognize <elake>. These symbols are identified by looking at the railroad diagram in figure 3. According to this diagram, the parser recognizes <elake> after the grammar symbols '{', ';', and block. In other words, when the parser recognizes these symbols, it can next recognize <elake>. We can see that the parser can next recognize block, ';', and '}' as well as <elake> when it recognizes '{', ';', or block. Therefore, the symbols

# Lake Symbols for Island Parsing

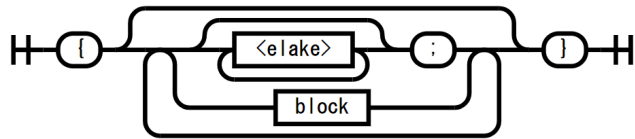

**Figure 3** Railroad diagram

block, ';', and '}' are the alternative symbols for <elake>. The alternative symbols are the destination symbols reached from the source symbols that we can also reach <elake> from.

Table 1 illustrates how algorithm 1 calculates $ALT(e_i)$ for every parsing expression $e_i$ in figure 2. To calculate the fixed point, the algorithm iterates five times. Each row represents a parsing expression $e_1$ to $e_{15}$. $BEGINNING(e_i)$ and $SUCCEED(e_i)$ are also shown for each $e_i$. Each iteration examines a parsing expression in the order from $e_1$ to $e_{15}$.

In the first iteration, for example, when the algorithm processes $e_3$, it updates $ALT(e_2)$ since $e_3$ is a zero-or-more expression $e_2*$. $ALT(e_3)$ is empty but $SUCCEED(e_3) = \{`\}`\}$. Thus, the updated $ALT(e_2)$ is $\{`\}`\}$. When the algorithm processes $e_9$, it updates $ALT(e_7)$ since $e_9$ is a prioritized-choice expression $e_7/e_8$. $ALT(e_9)$ is empty in this iteration and $BEGINNING(e_8) = \{block\}$. Thus, the updated $ALT(e_7)$ is $\{block\}$.

In the second iteration, when the algorithm processes $e_2$, it updates $ALT(e_9)$. Since $e_2$ is a nonterminal symbol stmt and its rule is stmt ← $e_9$, $ALT(e_2) = \{`\}`\}$ is added to $ALT(e_9)$. Thus, it updates $ALT(e_9)$ into $\{`\}`\}$. Then, when the algorithm processes $e_9$, it updates $ALT(e_7)$ again. Since $ALT(e_9)$ is already $\{`\}`\}$ at this time, $ALT(e_7)$ becomes $\{block, `\}`\}$.

In the final iteration, when the algorithm processes $e_{10}$, it updates $e_{15}$ into $ALT(e_{15}) \cup ALT(e_{10}) = \{`;`, block, `\}`\}$. This is what we need as the alternative symbols for <elake>.

Since we have obtained $ALT(e_{15})$, we can apply Step 2 in section 3.1 to the intermediate grammar shown in figure 2. The rule for <elake>:

<elake> <- !.

in the intermediate PEG is translated into the following rule:

<elake> <- !. / !(block / ';' / '}') .

in the normal PEG. This rule is semantically equivalent to the following rule:

<elake> <- !(block / ';' / '}') .

since !. does not match any input. Note that block, ';', and '}' are the alternative symbols for <elake>, which are the elements of $ALT(e_{15})$.

## 3.4 Limitations

The island parser looks ahead into the remaining text and attempts to match alternative symbols to check if the next token is water. The number of lookahead tokens is not limited and depends on what the alternative symbols are. However, at least one or more lookahead token is needed. In other words, our algorithm does not support the case that the alternate symbol for a lake symbol recognizes an empty input $\epsilon$. For example,

```
1  stmt <- expr ';'
2  expr <- <term> opt
3  opt <- '++'?
```

**Table 1** $ALT(e_i)$ for each iteration

| $e_i$ | operator | $BEGINNING(e_i)$ | $SUCCEED(e_i)$ | $ALT(e_i)$ for each iteration | | | | |
|---|---|---|---|---|---|---|---|---|
| | | | | 1 | 2 | 3 | 4 | 5 |
| $e_1$ | Terminal { | {'{'} | {stmt,'}'} | ∅ | ∅ | ∅ | ∅ | {'}'} |
| $e_2$ | Nonterminal stmt | {stmt} | {stmt,'}'} | {'}'} | {'}'} | {'}'} | {'}'} | {'}'} |
| $e_3$ | Zero-or-more $e_2*$ | {$\epsilon$, stmt} | {'}'} | ∅ | ∅ | ∅ | ∅ | ∅ |
| $e_4$ | Sequence $e_1 e_3$ | {'{'} | {'}'} | ∅ | ∅ | ∅ | {'}'} | {'}'} |
| $e_5$ | Terminal } | {'}'} | {stmt,'}'} | ∅ | ∅ | ∅ | ∅ | ∅ |
| $e_6$ | Sequence $e_4 e_5$ | {'{'} | {stmt,'}'} | ∅ | ∅ | {'}'} | {'}'} | {'}'} |
| $e_7$ | Nonterminal expr_stmt | {expr_stmt} | {stmt,'}'} | {block} | {block,'}'} | {block,'}'} | {block,'}'} | {block,'}'} |
| $e_8$ | Nonterminal block | {block} | {stmt,'}'} | ∅ | {'}'} | {'}'} | {'}'} | {'}'} |
| $e_9$ | Prioritized choice $e_7/e_8$ | {expr_stmt, block} | {stmt,'}'} | ∅ | {'}'} | {'}'} | {'}'} | {'}'} |
| $e_{10}$ | Lake <elake> | {<elake>, $\epsilon$} | {<elake>,';'} | {';'} | {';'} | {';', block} | {';', block,'}'} | {';', block,'}'} |
| $e_{11}$ | Zero-or-more $e_{10}*$ | {$\epsilon$, <elake>} | {';'} | ∅ | {block} | {block,'}'} | {block,'}'} | {block,'}'} |
| $e_{12}$ | Terminal ; | {';'} | {stmt,'}'} | ∅ | {block} | {block,'}'} | {block,'}'} | {block,'}'} |
| $e_{13}$ | Sequence $e_{11} e_{12}$ | {<elake>,';'} | {stmt,'}'} | ∅ | {block} | {block,'}'} | {block,'}'} | {block,'}'} |
| $e_{14}$ | Terminal . | {'.'} | ∅ | ∅ | ∅ | ∅ | ∅ | ∅ |
| $e_{15}$ | Not $!e_{14}$ | {$\epsilon$} | {<elake>,';'} | ∅ | {';'} | {';'} | {';', block} | {';', block,'}'} |

The ++ operator is optional. The alternate symbols for <term> are opt and ';' and the rule for <term> is:

```
<term> <- !(opt / ';') .
```

Since opt can recognize $\epsilon$, this *not*-predicate always fails. Thus, the lake symbol <term> could not recognize any symbol; it could not work as a wildcard-like symbol.

Our prototype system of the island-parser generator detects whether a lake symbol has $\epsilon$ as its alternative symbol. If our prototype detects such a lake symbol, it prints a warning message so that the user can modify the grammar. Our system reports that $\epsilon$ is the alternative symbol for a lake symbol <X> if there exists $Y \in ALT(<X>)$ such that $Y \leftarrow e_i$ and $\epsilon \in BEGINNING(e_i)$.

## 4 Experiments

We have implemented our prototype system *PEGIsland* in Python. PEGIsland is an island parser. It reads a grammar definition written in our extended PEG and then parses a given source program to build a parse tree based on that grammar. It performs packrat parsing [8], based on the internally generated normal PEG from the given grammar. The generated normal PEG can be optionally written out to a file. The parse tree is written out in the JSON format. Their intermediate nodes correspond to nonterminal symbols or lake symbols in the grammar. Their leaves correspond to terminal symbols.

We conducted experiments using PEGIsland to answer our research question, to what extent the lake symbols reduce the number of grammar rules for parsers. We implemented island parsers for Java and Python and compared the size of their grammars with the size of the full-featured grammars.

We also compared the size of the grammars written for island parsers with lakes and without lakes. Although our lake symbols help the definition of lakes, an island parser can be implemented without lakes. As we showed in section 2, writing a grammar without lakes is not difficult. It was easy when we only defined program_sea and program_water. It became difficult when we defined expr_sea and expr_water for a lake so that we could omit the definition of expr appearing in the middle of the island, if_else_stmt. We examined whether or not the use of the lakes effectively affected the number of grammar rules for island parsers. This was our second research question.

### 4.1 Java

For the full-featured Java grammar, we chose the grammar definition of an open-source PEG parser *MOUSE* [21]. It is based on the Java Language Specification for Java SE 8 Edition, with the corrections by the MOUSE developer to make it compatible with *javac*. The grammar consists of 279 PEG rules.

We implemented 36 island parsers on top of our PEGIsland parser. For each parser, we implemented two versions with lakes and without lakes. As an island, each of these parsers recognizes one of the nonterminal symbols in the full-featured grammar. According to the source-code comments, that grammar includes the nonterminal symbols corresponding to a program construct described in the dedicated section of the Java Language Specification [11]. We chose these nonterminal symbols as the islands. We implemented the island parsers so that the number of their grammar rules would be as small as possible. We used lake symbols to avoid defining irrelevant nonterminal symbols as far as the parser could correctly recognize an island. To verify that those island parsers correctly recognized their islands, we ran the island parsers to parse all the 7695 Java source files distributed as part of the Java 1.8 SDK. We then confirmed that these island parsers recognized the same number of islands as the number of the corresponding nonterminals that the full-featured Java parser recognized.

Table 2 lists the results. Each row represents the result of each island parser. For example, the second row represents that the parser recognizes `PackageDeclaration` as an island and it is described in Section 7.4 of the Java Language Specification (JLS). The number of the rules for its grammar is 22 and this grammar includes two lake symbols and three alternative symbols. # of *ALT* is the sum of the alternative symbols for every lake. We counted duplicate symbols more than once. The grammar defined without lakes needs 277 rules. `PackageDeclaration` is described in JLS 7.4 as follows:

```
1  PackageDeclaration:
2    {PackageModifier} package Identifier {. Identifier} ;
3  PackageModifier:
4    Annotation
```

It consists of an optional `PackageModifier`, the `package` keyword, comma-separated identifiers, and a semicolon. `PackageModifier` is an annotation. When we used a lake symbol, we could substitute a lake symbol for the identifiers between `package` and a semicolon. The alternate symbol for this lake symbol was a semicolon. We could also redefine `PackageModifier` as a lake symbol following '@'. However, when we did not use a lake symbol, we used only the `sea` symbol. Because we had to define the `Annotation` symbol for the `PackageModifier` symbol, we had to define almost all the other nonterminals included in the full-featured grammar. An annotation may include a lambda expression and its body may include almost all kinds of statements and declarations. The number of the grammar rules in this case was 277; only two rules could be omitted.

Figure 4 illustrates the comparison of the number of rules listed in table 2. The height of bars indicates the number of rules in percentage. The number of the rules for the full-featured grammar is 100 %. Each pair of bars corresponds to the island parsers for recognizing the same nonterminal symbol. The left bar indicates the island parser without lakes while the right bar indicates with lakes.

**Table 2** The island parsers for Java

| Section | Nonterminal symbol | # of grammar rules without lakes | # of grammar rules with lakes | # of lakes | # of ALT |
|---|---|---:|---:|---:|---:|
| 7.3 | CompilationUnit | 279 | 16 | 1 | 1 |
| 7.4 | PackageDeclaration | 277 | 22 | 2 | 3 |
| 7.5 | ImportDeclaration | 280 | 137 | 5 | 8 |
| 7.6 | TypeDeclaration | 280 | 109 | 6 | 8 |
| 8.1 | ClassDeclaration | 274 | 80 | 5 | 6 |
| 8.3 | FieldDeclaration | 274 | 206 | 3 | 3 |
| 8.4 | MethodDeclaration | 274 | 130 | 5 | 8 |
| 8.6 | InstanceInitializer | 274 | 274 | 0 | 0 |
| 8.7 | StaticInitializer | 274 | 21 | 1 | 1 |
| 8.8 | ConstructorDeclaration | 274 | 236 | 1 | 1 |
| 8.9 | EnumDeclaration | 274 | 94 | 4 | 7 |
| 9.1 | InterfaceDeclaration | 274 | 88 | 4 | 6 |
| 9.3 | ConstantDeclaration | 274 | 204 | 3 | 3 |
| 9.4 | InterfaceMethodDeclaration | 274 | 133 | 5 | 8 |
| 9.6 | AnnotationTypeDeclaration | 274 | 39 | 2 | 2 |
| 9.7 | Annotation | 274 | 26 | 1 | 1 |
| 10.6 | ArrayInitializer | 274 | 274 | 0 | 0 |
| 14.2 | Block | 274 | 274 | 0 | 0 |
| 14.4 | LocalVariableDeclarationStatement | 274 | 274 | 0 | 0 |
| 14.5 | Statement | 274 | 274 | 0 | 0 |
| 14.8 | StatementExpression | 274 | 274 | 0 | 0 |
| 14.11 | SwitchBlock | 274 | 23 | 1 | 1 |
| 14.14 | BasicForStatement | 274 | 231 | 4 | 13 |
| 14.20 | TryStatement | 274 | 29 | 2 | 2 |
| 15.2 | Expression | 274 | 274 | 0 | 0 |
| 15.8 | Primary | 274 | 274 | 0 | 0 |
| 15.9 | ClassCreator | 274 | 72 | 3 | 6 |
| 15.10 | ArrayCreator | 274 | 73 | 2 | 6 |
| 15.12 | Arguments | 274 | 274 | 0 | 0 |
| 15.15 | UnaryExpression | 274 | 274 | 0 | 0 |
| 15.16 | CastExpression | 274 | 274 | 0 | 0 |
| 15.17-24 | InfixExpression | 274 | 274 | 0 | 0 |
| 15.25 | ConditionalExpression | 274 | 274 | 0 | 0 |
| 15.26 | AssignmentExpression | 274 | 274 | 0 | 0 |
| 15.27 | LambdaExpression | 274 | 82 | 2 | 8 |
| 15.28 | ConstantExpression | 274 | 174 | 1 | 3 |

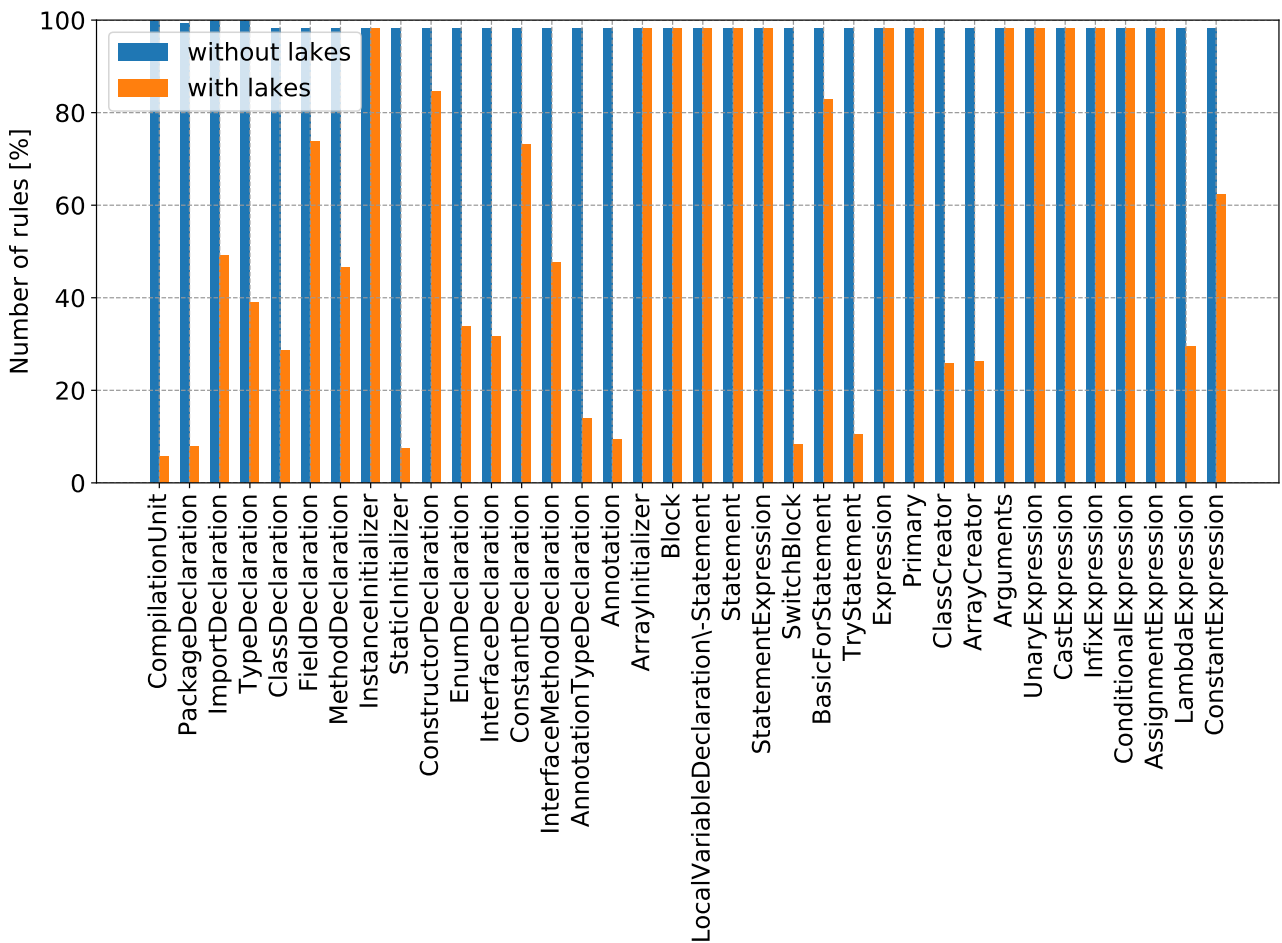

**Figure 4** The number of the rules for each island parser for Java

**Without Lakes**

The experiments revealed that an island parser for Java did not reduce the number of grammar rules unless it uses lakes. The best case was 274 rules whereas the full-featured Java parser needs 279 rules; only 2 % reduction. In Java, most program constructs may include almost all kinds of program constructs. As shown above, a PackageDeclaration may include an annotation and an annotation may include a lambda expression. A lambda expression may include all kinds of statements and declarations in its body. A PackageDeclaration, therefore, may indirectly include all kinds of statements and declarations. Without a lake, the grammar for the island parser for PackageDeclaration must include the rules for almost all nonterminal symbols in Java.

**With Lakes**

If lakes are used, 22 of 36 island parsers needed a smaller number of grammar rules than the full-featured Java parser and the island parser using no lakes. The 22 island parsers are 61 % among all. They needed 16 to 231 rules. CompilationUnit needed

**Lake Symbols for Island Parsing**

16 rules and it achieved 94 % reduction. `BasicForStatement` needed 231 rules and it achieved 17 % reduction. The rules included 1 to 6 lake symbols. The lake symbols automated the enumeration of 1 to 13 alternative symbols. They enumerated 1.7 alternative symbols per lake.

Some island parsers needed a number of grammar rules for correctly recognizing an island. For example, `MethodDeclaration` needed 130 rules despite 5 lake symbols included in the rules. It needed such a large number of rules because it had to be distinguished from `InterfaceMethodDeclaration`. The following code snippet:

```
abstract void foo(int x);
```

is recognized as a `MethodDeclaration` when it is enclosed in a class declaration. However, it is recognized as an `InterfaceMethodDeclaration` when it is enclosed in an interface declaration. To recognize only a `MethodDeclaration`, the island parser had to also recognize `ClassDeclaration` and `InterfaceDeclaration` so that it could distinguish `MethodDeclaration` from `InterfaceMethodDeclaration`.

`Block` needed 274 rules for the same reason. Its grammar is just slightly smaller than the grammar for the full-featured Java parser. A `Block` recognizes a token sequence that starts with { and ends with } but other nonterminal symbols also recognize such a sequence. `ClassBody`, `SwitchBlock`, `EnumBody`, and so on recognize the sequence. To distinguish them, we had to define the rules for recognizing all of them. If we did not distinguish them but we just wanted to recognize any kind of block-like structure, the grammar for this island parser would be much smaller.

The island parsers for expressions, such as `Expression`, `Primary`, and `UnaryExpression`, needed a large number of rules despite the use of lakes. Since Java supports infix operators, these nonterminal symbols do not start with a particular keyword. It is difficult to spot where these symbols start in the sea or a lake. Thus, we had to recognize all the program constructs that may enclose such an expression. This is why the grammars had to include almost all the nonterminal symbols that the full-featured grammar does. Although the island parser for a most kind of expression needed a large grammar, `LambdaExpression` was an exception. It only needed 82 rules since it always starts with a parenthesized parameter list or comma-separated identifiers, and an arrow ->.

### 4.2 Python

For the full-featured Python grammar, we chose the grammar distributed with Python 3.7.4. Since it was a LL(1) grammar, we manually translated it into a PEG. The number of nonterminal symbols in the grammar was 187. The original grammar requires that the input text is preprocessed, so that its indentations will be converted into tokens `INDENT` and `DEDENT`. Our PEG translation of this grammar also takes similarly preprocessed text as its input.

We implemented 20 island parsers on top of our PEGIsland parser. For each parser, we implemented two versions with lakes and without lakes. As an island, each of these parsers recognizes one of the nonterminal symbols in the full-featured grammar. We selected 20 nonterminal symbols from the full-featured grammar. These 20 symbols correspond to the program constructs described in a dedicated section 7 or 8 in the Python Language Reference [20]. We excluded expression statements and assignment statements but added lambda expressions from Section 6 because we knew that island parsing was not suitable for expressions after we had conducted the experiments for Java. To verify the correctness of our island parsers, we ran the island parsers to parse 1634 Python source files under the lib directory of the Python 3.7.4 distribution.

**Table 3** The island parsers for Python

| Section | Nonterminal symbol | # of grammar rules without lakes | # of grammar rules with lakes | # of lakes | # of ALT |
|---|---|---:|---:|---:|---:|
| 6.13. | lambdef | 105 | 20 | 3 | 8 |
| 7.3. | assert_stmt | 107 | 12 | 1 | 2 |
| 7.4. | pass_stmt | 8 | 8 | 0 | 0 |
| 7.5. | del_stmt | 107 | 12 | 1 | 2 |
| 7.6. | return_stmt | 107 | 12 | 1 | 2 |
| 7.7. | yield_stmt | 111 | 16 | 1 | 2 |
| 7.8. | raise_stmt | 107 | 12 | 1 | 2 |
| 7.9. | break_stmt | 8 | 8 | 0 | 0 |
| 7.10. | continue_stmt | 8 | 8 | 0 | 0 |
| 7.11. | import_stmt | 25 | 17 | 3 | 5 |
| 7.12. | global_stmt | 14 | 12 | 1 | 2 |
| 7.13. | nonlocal_stmt | 14 | 12 | 1 | 2 |
| 8.1. | if_stmt | 188 | 34 | 5 | 13 |
| 8.2. | while_stmt | 188 | 31 | 5 | 12 |
| 8.3. | for_stmt | 188 | 36 | 6 | 15 |
| 8.4. | try_stmt | 188 | 37 | 5 | 13 |
| 8.5. | with_stmt | 188 | 33 | 6 | 15 |
| 8.6. | funcdef | 188 | 36 | 6 | 13 |
| 8.7. | classdef | 188 | 35 | 6 | 14 |
| 8.8. | async_funcdef | 188 | 41 | 5 | 12 |

Table 3 lists the results. Each row represents the result of each island parser. Figure 5 illustrates the comparison of the number of rules listed in table 3.

**Without Lakes**

As in Java, 8 island parsers could not reduce the number of grammar rules without lakes. For example, if_stmt and while_stmt could not reduce because if and while statements may enclose all other program constructs in its body. They rather slightly increased the number of rules due to the introduction of the *sea* symbol.

However, because the Python grammar is relatively simpler than Java's, the other 12 island parsers implemented without lakes could successfully reduce the number of grammar rules. For example, pass_stmt, break_stmt, and continue_stmt needed only 8 rules. They achieved 96 % reduction. The 6 island parsers such as lambdef and assert_stmt achieved 41 % to 44 % reduction.

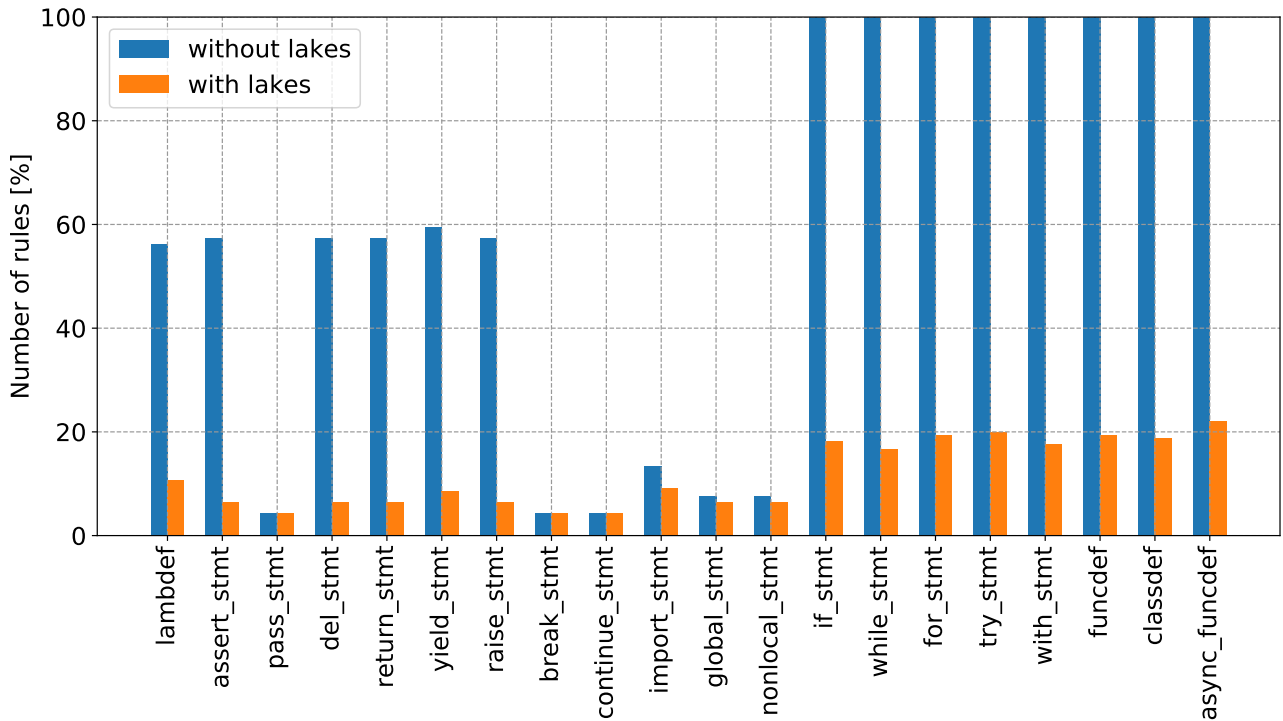

**Figure 5** The number of the rules for each islands parsing for Python

**With Lakes**
All the island parsers implemented with lakes successfully reduced the number of grammar rules. They achieved more than 78 % reduction. Unlike Block in Java, different nonterminal symbols in Python never recognize the same token sequence as their parts. This was a reason why lake symbols effectively reduced the number of grammar rules. Every lake symbol automated the enumeration of 2.4 alternative symbols in average.

Compared to the island parser without lakes, the island parser with lakes achieved larger reduction of the grammar rules. The lake symbols could reduce the number of grammar rules even when the island parsers without lakes could reduce it against the full-featured parser. For example, the island parser for lambdef without lakes reduced 44 % against the full-featured parser but the parser with lakes reduced 89 %. This was 45 % improvement. The exceptions were the three parsers pass_stmt, break_stmt, and continue_stmt, which needed only 8 rules. They could not reduce the number of their grammar rules against the parser without lakes.

### 4.3 Summary of the Experiments

For our first research question, to what extent the lake symbols reduce the number of grammar rules for parsers, the results of our experiments revealed that the lake symbols effectively reduced the number of grammar rules for Java and Python except the nonterminal symbols for expressions and some others. The lake symbols worked better for Python. Excluding the nonterminal symbols for expressions (and lambdef), the average reduction rate for Python was 89 % whereas it was 42 % for Java. This would be related to the syntactical complexity of the two languages. Python is much simpler than Java. However, in both languages, the lake symbols were not effective when the start and the end of the nonterminal symbol were not spot by a particular symbol. They were not effective either when more than one nonterminal symbols recognized a token sequence as their parts.

For our second research question, whether the use and disuse of lakes affect the number of grammar rules for island parsing, the results of the experiments revealed that the use of lakes was indispensable to reduce the number of grammar rules. However, in several cases, the use of lakes did not affect the number of grammar rules. The number of rules did not change no matter if lakes are used or not.

One of the threats of validity is that we used only one grammar as the full-featured grammar for each language, Java and Python. If we use a different full-featured grammar, we might see different results. Another threat is that we only examined Java and Python. The effects of the lakes may change depending on the syntactical complexity of the language.

## 5 Related Work

Koopa[1] is a parser generator with support for island grammars that has been used to implement an industrial-strength Cobol parser. Koopa provides the skip-to operator to support the description of *water*. The skip-to operator skips anything it encounters up to the given pattern. While the skip-to operator cannot handle nested islands in *water*, our lake symbol can handle them.

The semi-parsing approach used for Agile parsing [6] can be considered as a variant of island parsing. It proposed the use of the not operator in TXL [5] so that the user can manually specify the alternative symbols, which our lake symbol automatically derives from the rest of the grammar.

Bounded seas [15] automatically calculate the alternative symbols and use them for the not predicate as our lake symbols do. However, as far as we understand, the bounded seas calculate only a subset of all the alternative symbols, which corresponds to our SUCCEED. Thus, its applicability is limited.

The generalized parsers, such as CYK [27], GLL [23], and GLR [25], can be used for island parsing although they deal with context-free grammars (CFGs) but our work deals with PEGs. The generalized parsers generate not a parse tree but parse forest, which consists of several candidates of parse tree for a given source program. This is useful to recognize *water* without specifying complex alternative symbols. However, the user must manually disambiguate the forest to obtain a single parse tree.

---

[1] https://github.com/krisds/koopa, last accessed 2020-10-01.

**Lake Symbols for Island Parsing**

Afroozeh, Bach, Van den Brand, Johnstone, Manders, Moreau, and Scott [1] addressed this disambiguation for their GLL-based island parsing used for the Tom language. They proposed to use pattern matching for the disambiguation.

Lavie and Tomita [16] proposed the GLR* parser, which can parse any input sentence by ignoring unrecognizable parts of the sentence. The application domain of GLR* is speech recognition. GLR* supports the *sea* but not the *lakes*.

Goloveshkin and Mikhalkovich [10] proposed the special terminal symbol Any that matches zero or more tokens constituting uninteresting parts of a given input program. Any can be used to define the *sea* but not the *lakes*.

Fuzzy parsing is the parsing approach where only certain parts of a programming language is recognized. A hand-written fuzzy parser was presented for C++ in [4]. A framework for fuzzy parser was proposed by Koppler [14]. In the framework, the scanner searches an input source program to spot the anchors where the parser starts parsing. The user must manually implement that scanner.

Klusener and Lämmel [13] proposed the method that constructs a skeleton grammar for tolerant parsing from a given grammar. The parsers derived from the skeleton grammar can be considered as an island parser. Their proposal does not support lakes but could be extended to support our lake symbols.

## 6 Conclusion and Future Work

This paper presented the lake symbols for island parsing and an extended PEG supporting lake symbols. The lake symbol is a special grammatical symbol like a wildcard. The user can use lake symbols to define rules for an island with lakes. The user can also define a rule for each lake symbol to specify inner islands or recursive structures inside the lake. This paper proposed an algorithm that translates our extended PEG to the normal PEG that can be given to existing PEG parsers. We experimentally revealed that lake symbols effectively reduced the size of the grammars for island parsers for Java and Python, compared to their fully detailed grammars.

### 6.1 Future Work

We have defined the semantics of lake symbols by translating them into a specific grammar. Defining its semantics without the translation is our future work.

The implementation of lake symbols for another grammar class is also future work. The concept of lake symbols will be applicable not only PEGs but also context-free grammars (CFGs) and their subsets such as LR and LL grammars. An issue is the ambiguity of a grammar when introducing lake symbols into another grammar class than PEGs. Since a lake symbol works as a wildcard, the grammar written with lake symbols tends to be ambiguous. This ambiguity causes shift/reduce conflicts, for example, when using the LR parsing. We expect that the alternative symbols proposed in this paper can be used for disambiguating the grammar. The algorithm for calculating the alternative symbols in CFGs would be similar to what we presented in this paper. In this paper, we selected PEGs since PEG's prioritized choice operator does not cause ambiguity even when lake symbols are used.

Other future work is to evaluate with more practical applications. Our experiment confirmed that the number of rules to extract a specific kind of programming construct decreased with lakes. Although we believe that reducing rules ease the construction of island parsing, the relation between the real workload and the number of rules has not been studied experimentally. Besides, while we assumed that the application is interested in only one kind of programming constructs in our experiment, a realistic application may need more than one kind of programming constructs.

## A  Parsing Expression Grammars

This appendix gives a brief explanation on PEGs used in this paper. According to the literature [9], a parsing expression grammar (PEG) is a 4-tuple $G = (V_N, V_T, R, e_s)$, where $V_N$ is a finite set of non-terminal symbols, $V_T$ is a finite set of terminal symbols, $R$ is a finite set of rules, $e_s$ is a start expression, and $V_N \cap V_T = \emptyset$. Each rule $r \in R$ is represented as $A \leftarrow e$, where $A$ is a non-terminal symbol and $e$ is a parsing expression.

A parsing expression is similar to a regular expression in that it presents a pattern of strings to be recognized. A parsing expression consists of a sequence of terminal/non-terminal symbols and operators. The operators are summarized in table 4. The semantics of operators except /, !, and & are the same as those in regular expressions. / and ! have important roles in island parsing.

/ operator is the prioritized choice operator. When both the left operand $e_1$ and right operand $e_2$ match an input string, $e_1$ is always prioritized. Hence, ambiguity does not exist in a parsing expression. In island parsing, we use a wildcard to skip water. If we use a wildcard as one of the operands of a ordinal choice operator, the grammar becomes ambiguous. In PEG, the user can disambiguate this situation by putting the wildcard on the right hand of / operator.

! is the lookahead not-predicate. The parser is expected to lookahead the input to check if the operand of ! recognize the head of the input without consuming any character.

▪ **Table 4**  Operators for parsing expressions

| Operator | Precedence | Description |
|:---:|:---:|:---:|
| . | 5 | Any character |
| $(e)$ | 5 | Grouping |
| $e?$ | 4 | Optional |
| $e*$ | 4 | Zero-or-more |
| $!e$ | 3 | Not-predicate (Negative lookahead) |
| $\&e$ | 3 | And-predicate (Positive lookahead) |
| $e_1 e_2$ | 2 | Sequence |
| $e_1/e_2$ | 1 | Prioritized choice |

## B Fixed-point Constraints

This appendix shows the constraints satisfied by the fixed-points computed by Algorithms 1–3.

### B.1 ALT

The *ALT* sets can be computed as the fixed-point over the following constraints:

1. If $e_i$ is a nonterminal or lake symbol $\mathscr{S}$ and $\mathscr{S} \leftarrow e_j \subseteq R'$, then $ALT(e_i) \in ALT(e_j)$
2. If $e_i$ is $e_j*$, $e_j+$, or $e_j?$, then $ALT(e_j) = ALT(e_i) \cup SUCCEED(e_i)$
3. If $e_i$ is $!e_j$, then $ALT(e_j) = SUCCEED(e_i)$
4. If $e_i$ is $\&e_j$, then $ALT(e_j) = ALT(e_i)$
5. If $e_i$ is $e_j/e_k$, then $ALT(e_k) = ALT(e_i)$ and
    a. If $\epsilon \in BEGINNING(e_k)$,
       then $ALT(e_j) = ALT(e_i) \cup (BEGINNING(e_k) - \{\epsilon\}) \cup SUCCEED(e_k)$
    b. Else $ALT(e_j) = ALT(e_i) \cup BEGINNING(e_k)$
6. If $e_i$ is $e_j e_k$, then $ALT(e_j) = ALT(e_i)$ and
    a. If $\epsilon \in BEGINNING(e_j)$, then $ALT(e_k) = ALT(e_i)$
    b. Else $ALT(e_k) = \emptyset$

### B.2 BEGINNING

The *BEGINNING* sets can be computed as the fixed-point over the following constraints:

1. If $e_i$ is a terminal symbol $\alpha$, then $BEGINNING(e_i) = \{\alpha\}$
2. If $e_i$ is a nonterminal or lake symbol $\mathscr{S}$ and $\mathscr{S} \leftarrow e_j \in R'$, then $BEGINNING(e_i) = \{\mathscr{S}\}$
3. If $e_i$ is $e_j?$ or $e_j*$, then $BEGINNING(e_i) = BEGINNING(e_j) \cup \{\epsilon\}$
4. If $e_i$ is $e_j+$, then $BEGINNING(e_i) = BEGINNING(e_j)$
5. If $e_i$ is $!ej$ or $\&e_j$, then $BEGINNING(e_i) = \{\epsilon\}$
6. If $e_i$ is $e_j/e_k$, then $BEGINNING(e_i) = BEGINNING(e_j) \cup BEGINNING(e_k)$
7. If $e_i$ is $e_j e_k$
    a. If $\epsilon \in BEGINNING(e_j)$,
       then $BEGINNING(e_i) = (BEGINNING(e_j) - \{\epsilon\}) \cup BEGINNING(e_k)$
    b. Else $BEGINNING(e_i) = BEGINNING(e_j)$

### B.3 SUCCEED

The *SUCCEED* sets can be computed as the fixed-point over the following constraints:

1. If $e_i$ is a nonterminal or lake symbol $\mathscr{S}$ and $\mathscr{S} \leftarrow e_j \in R'$, then $SUCCEED(e_i) \subseteq SUCCEED(e_j)$
2. If $e_i$ is $e_j?$, then $SUCCEED(e_j) = SUCCEED(e_i)$

3. If $e_i$ is $e_j*$ or $e_j+$, then $SUCCEED(e_j) = SUCCEED(e_i) \cup BEGINNING(e_i) - \{\epsilon\}$
4. If $e_i$ is $!e_j$ or $\&e_j$, then $SUCCEED(e_j) = \emptyset$
5. If $e_i$ is $e_j/e_k$, then $SUCCEED(e_j) = SUCCEED(e_i)$ and $SUCCEED(e_k) = SUCCEED(e_i)$
6. If $e_i$ is $e_j e_k$, then $SUCCEED(e_k) = SUCCEED(e_i)$
   a. If $\epsilon \in BEGINNING(e_k)$,
      then $SUCCEED(e_j) = (BEGINNING(e_k) - \{\epsilon\}) \cup SUCCEED(e_k)$
   b. Else $SUCCEED(e_j) = BEGINNING(e_k)$

**About the authors**

**Katsumi Okuda** is a Ph.D. student at The University of Tokyo. He is also a head researcher at Mitsubishi Electric Corporation. His research interests include programming languages, generative programming, and their application to embedded systems. Contact him at Okuda.Katsumi@eb.MitsubishiElectric.co.jp.

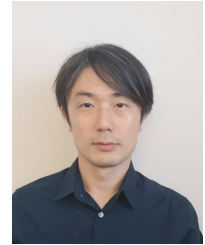

**Shigeru Chiba** is Professor at The University of Tokyo. Shigeru's research focuses on programming language design, implementation, tools, and libraries. He has been publishing papers on reflection and meta-programming, aspect-oriented programming, and embedded domain-specific languages at OOPSLA, ECOOP, and other conferences. He has also served as a program committee member at those conferences. He received the 2012 IBM Faculty Award. He is also a primary developer of Javassist, which is a Java bytecode engineering toolkit widely used in industry and academia. He received his Ph.D. in computer science from the University of Tokyo. Contact him at chiba@acm.org

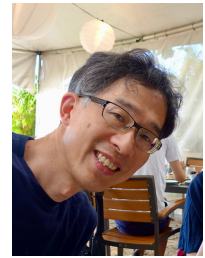